\documentclass[aps,pra,reprint,groupedaddress,showpacs]{revtex4-1}
\usepackage{bm}
\usepackage{graphicx}
\usepackage{times}
\usepackage{color}
\usepackage{mathtools}

\begin{document}

\title{Dynamical instability in the $S=1$ Bose-Hubbard model}

\author{Rui \surname{Asaoka}$^{1}$\email[]{asaoka@olive.apph.tohoku.ac.jp}, 
Hiroki \surname{Tsuchiura$^{1}$},
Makoto \surname{Yamashita}$^{2}$, and 
Yuta \surname{Toga}$^{1}$}

\affiliation{$^{1}$Department of Applied Physics, Tohoku University, Sendai 980-8579, Japan \\
$^{2}$NTT Basic Research Laboratories, NTT Corporation, Atsugi, Kanagawa 243-0198, Japan \\
}

\date{\today}

\begin{abstract}
We study the dynamical instabilities of superfluid flows in the $S=1$ Bose-Hubbard model. The time evolution of each spin component in a condensate is calculated based on the dynamical Gutzwiller approximation for a wide range of interactions, from a weakly correlated regime to a strongly correlated regime near the Mott-insulator transition. Owing to the spin-dependent interactions, the superfluid flow of the spin-1 condensate decays at a different critical momentum from a spinless case when the interaction strength is the same. We furthermore calculate the dynamical phase diagram of this model and clarify that the obtained phase boundary has very different features depending on whether the average number of particles per site is even or odd.
Finally, we analyze the density and spin modulations that appear in association with the dynamical instability. We find that spin modulations are highly sensitive to the presence of a uniform magnetic field. 
\end{abstract}

\pacs{03.75.Kk, 03.75.Mn}

\maketitle

\section{Introduction}

A recent experimental development in optical lattices offers the unprecedented potential to study the dynamical properties of many-body interacting ultracold atoms \cite{bloch,morsch}.  
In particular, the superfluid flow of a Bose-Einstein condensate (BEC) loaded on a lattice exhibits a novel instability called dynamical instability that was predicted a decade ago \cite{wu_niu,wu_niu03},  which has attracted
much attention both theoretically \cite{smerzi,modugno,altman,polkov} and experimentally \cite{raman,cataliotti,fallani,mun,ferris}. 
The dynamical instability is induced by the interplay between the lattice periodicity and nonlinearity due to the inter-particle interactions in the BEC. When the system becomes 
dynamically unstable, the energy of an excited mode has an imaginary part \cite{wu_niu03}. Therefore, an arbitrary small density fluctuation in a uniform superfluid flow grows exponentially in time, resulting in a drastic decay of the original flow.  These features are in contrast with the well-known Landau instability, which is the energetic instability caused by decaying from the initial metastable state.

Dynamical instability itself is widely seen in various nonlinear systems governed by classical fluid mechanics. However, using ultracold atoms, we can now advance the study of superfluid instabilities to the next stage, namely  dynamical instabilities in systems with internal degrees of freedom. 
It has been known that multicomponent systems exhibit rich physics such as diverse quantum phases in an optical lattice \cite{altman_multi,kuklov,catani,buchler,gunter,best,sugawa}, and the dynamical instability of multicomponent bosons has also recently been studied \cite{hui,hooley,ruostekoski,wernsdorfer}. Moreover, bosons with unfrozen spin degrees of freedom specifically exhibit complex and intriguing phenomena caused by spin mixing processes \cite{stamper}.
The spin-1 bosons have therefore been investigated intensively as the simplest bosonic system with unfrozen spin degrees of freedom.
A series of studies have revealed interesting instabilities in the spin-1 BEC based on the Gross-Pitaevskii equation, spin mixing instability \cite{pu,chang}, spin counterflow instability \cite{law2,fujimoto}, and the  spontaneous formation of spin domains \cite{zhang,zhang2}. These phenomena are specific to the spin-1 bosonic system and have hardly been understood only by conventional linear stability analysis \cite{hui,wernsdorfer} because of the spin mixing process inherent in the system.

The spin-1 bosons in optical lattices are well described by the $S=1$ Bose-Hubbard model (BHM)  \cite{law,demler,imambe}.
The phase diagram and the static properties of this model have already been extensively studied using several theoretical methods \cite{krut,tsuchiya,kimura,yama,rizzi,sara,apaja,batrouni,toga}.
From these studies, the Gutzwiller-type variational wave functions are good at capturing the superfluid (SF) to Mott-insulator (MI) transition in the $S=1$ BHM, aside from spin correlations in the MI phase \cite{toga,parny,parny2}. It has been found that the SF-MI transition in this model strongly depends on whether the average particle number per site, $n$, is even or odd \cite{tsuchiya,kimura}.

We study the effect of spin interaction in the dynamical instability according to the following three interests.
First, whether does the parity about the average number of particles per site as mentioned above also appear in the dynamical phase diagram or not? This motivates us to explore a role of spin degrees of freedom in dynamical phenomena of a superfluid, which remains to be clarified.
Second, how do spin mixing processes among spin components in spin-1 superfluid flows affect the dynamical instability? Spin mixing, which is an important feature in bosonic spin systems, does not exist in classical fluids and multicomponent cold atom systems. Therefore, the effect of spin mixing on instabilities of fluids itself is intriguing.
Finally, we are interested in the very recent development of experimental techniques for observing spin dynamics of condensates in optical lattices as reported by L. Zhao {\it et al.}\,\cite{zhao}. It was revealed experimentally that the intensity of lattice potential significantly affects spin mixing dynamics in a spin-1 system. The experiment of the dynamical instability in a spin-1 system is therefore expected to be demonstrated in the near future.

In this paper, we analyze the dynamical instability of the spin-1 condensate in the $S=1$ BHM for a wide range of interaction parameters with antiferromagnetic or ferromagnetic interactions, focusing on the stability of spin-resolved superfluid flows. 
First, we reveal how the spin mixing process affects the real-time evolution of each spin component in the flow. We employ the dynamical Gutzwiller approximation that was used by Altman {\it et al.} \cite{altman,polkov} to analyze the dynamical instability of an SF in the spinless BHM. Recently, Natu {\it et al.} also applied this method to the $S=1$ BHM and they calculated the low-lying excitation spectrum \cite{natu}.
We show the dynamical decay of the $S=1$ superfluid flow and the corresponding time development of the spin components.
Next, we demonstrate the parity dependence of dynamical instability in the $S=1$ BHM constructing dynamical phase diagrams. In the antiferromagnetic case, the stable flow region on the phase diagram shrinks when the average number of particles is odd, while it grows for the even average numbers compared with the spinless case. We find that this phenomenon is caused by the spin mixing process.
Finally, we discuss the density and spin modulations associated with the dynamical instability with or without a uniform magnetic field.

\section{Model and Method}\label{sec:MM}

The Hamiltonian of the $S=1$ BHM is given as \cite{demler}
%
\begin{eqnarray}
 {\cal H} &=& -t\sum_{\langle i,j\rangle}\sum_{\gamma} 
\left( \hat{a}_{i,\gamma}^{\dagger}\hat{a}_{j,\gamma} 
     + \hat{a}_{j,\gamma}^{\dagger}\hat{a}_{i,\gamma} \right)
                  - \mu\sum_{i}\hat{n}_{i}
\nonumber \\
          & &+ \frac{U_{0}}{2}\sum_{i}\hat{n}_{i}(\hat{n}_{i}-1) 
             + \frac{U_{2}}{2}\sum_{i}\left( \hat{\bm S}_{i}^{2} 
                                             - 2\hat{n}_{i} \right) ,
\label{hamil}
\end{eqnarray}
where $t$ is the hopping amplitude of bosons, $\langle i,j \rangle$ in the summation denotes the pairs of nearest neighbors, $\mu$ is the chemical potential,
$U_{0} (>0)$ is the on-site spin-independent repulsion,
and $U_{2}$ is the on-site spin-dependent interaction.
In cold atom systems, the $U_{2}$ value depends on the $s$-wave scattering length, which is specific to atom species; for example, $U_{2}>0$ ($<0$) for Na (Rb) atoms.
$\hat a_{i,\gamma}$ is the annihilation operator of a boson at site $i$ with a spin state $\gamma$ ($=0, \pm 1$),
the local particle number operator $\hat n_{i} = \sum_\gamma \hat{a}_{i,\gamma}^{\dagger}\hat{a}_{i,\gamma}$,
and $\hat{\bm S}_{i}=\sum_{\gamma,\gamma'}\hat{a}_{i,\gamma}^{\dagger}{\bm S}_{\gamma,\gamma'}\hat{a}_{i,\gamma'}$ is the spin operator at site $i$ where ${\bm S}_{\gamma,\gamma'}$ corresponds to the spin-1 matrices.
The square of the local spin operator $\hat{\bm S}_{i}^{2}$ is represented in a more convenient formula: $\hat{\bm S}_{i}^{2}=(\hat{S}_{i,-}\hat{S}_{i,+}+\hat{S}_{i,+}\hat{S}_{i,-})/2+\hat{S}^{2}_{i,z}$ where the ladder operators are defined by $\hat{S}_{i,-}=\sqrt{2}(\hat{a}_{i,-1}^{\dagger}\hat{a}_{i,0}+\hat{a}_{i,0}^{\dagger}\hat{a}_{i,1})$ and $\hat{S}_{i,+}=\hat{S}_{i,-}^{\dagger}$, correspondingly.
This formula can also be written in terms of creation and annihilation operators,
\begin{eqnarray}
\hat{\bm S}_{i}^{2}&=&(\hat{n}_{i,1}-\hat{n}_{i,-1})^{2}
                      +\hat{n}_{i}+\hat{n}_{i,0}+2\hat{n}_{i,1}\hat{n}_{i,0}+2\hat{n}_{i,0}\hat{n}_{i,-1}
\nonumber \\
                        & &+2\hat{a}_{i,1}^{\dagger}\hat{a}_{i,-1}^{\dagger}(\hat{a}_{i,0})^{2}
                             +2(\hat{a}_{i,0}^{\dagger})^{2}\hat{a}_{i,1}\hat{a}_{i,-1}     .
\label{sp}
\end{eqnarray} 
The last two terms in Eq.\,(\ref{sp}) induce spin mixing between the $S_z=\pm 1$ and $S_z=0$ states, which enriches the physics of this model compared with spinless models or multi-component models without any mixing of components.

We first investigate the quantum dynamics of this model within the dynamical Gutzwiller scheme \cite{altman,polkov}. 
The variational wave function for the $S=1$ BHM can be written as the direct product of superposition states
at each lattice site
\begin{equation}
|\Psi_{\mathrm G}\rangle = \prod_{i}\left[
\sum_{\substack{n_{i, 0} \\ n_{i, \pm 1}}}
 f_{i}(n_{i,1},n_{i,0},n_{i,-1}) | n_{i,1}, n_{i,0}, n_{i,-1} \rangle
 \right]   ,
\label{GWF}
\end{equation}
where $| n_{i,1}, n_{i,0}, n_{i,-1} \rangle$ denotes the local Fock state determined by the local number of atoms for each spin component at site $i$. Here the Gutzwiller parameters are normalized as $\sum_{\gamma,n_{i,\gamma}}|f_{i}(n_{i,1},n_{i,0},n_{i,-1})|^{2}=1$. 
Minimizing $\langle \Psi_{\mathrm G}| i\hbar\frac{\partial}{\partial t} - {\cal H} |\Psi_{\mathrm G}\rangle$
on the basis of the time-dependent variational principle, we derive equations of motion with respect to these Gutzwiller parameters \cite{natu}. 
The equations are explicitly shown in the Appendix. Note that $p(t)$ in Eq.\,(\ref{eom2}) corresponds to the relative momentum between a condensate and a lattice for a condensate on a moving lattice or a moving condensate on a stationary lattice. We introduce $p(t)$ as the phase difference between particles at adjacent sites using the transformation: $a_{j,\gamma}\mapsto a_{j,\gamma}\mathrm{e}^{ip(t)j}$ (note that, $t$ represents time here).
In the time-evolution calculations, we assume $p=0$ at the initial time and the system stays in the ground state initially for given $U_{0}$ and $U_{2}$.  The momentum is then increased linearly with time at the acceleration rate $\alpha$: $p(t)=\alpha t$. 
We perform this procedure almost adiabatically by choosing a very small rate $\alpha = 0.005$. 
Since loss of atoms is neglected in our study,  the total number of particles should be conserved during time evolution. We ensure the number conservation from the fact that the filling $n=n_{1}+n_{0}+n_{-1}$ (i.e., the average particle number per site) is kept constant within the numerical precision.
%
The calculated system is a two-dimensional lattice with a unit size $L=40\times 2$ with periodic boundary conditions, and we set the hopping amplitude $t=1$ as a unit of energy. In our calculation, the sum of the wavefunction (\ref{GWF}) is limited to a finite number of states to reduce the number of computational tasks. We confirmed  that the truncation does not produce any noticeable differences in the numerical results.

In our calculations, velocity of a superfluid flow becomes quantized owing to periodic boundary conditions. However, the decay of a flow from the initial state to the lower winding number states does not occur even in a ring geometry because of the conditions we assume here, i.e., at zero temperature without any thermal fluctuations. Our system therefore essentially becomes equivalent to the non-periodic systems that are generally realized in the optical lattice experiments to observe the dynamical instability. Actually, Mun {\it et al.}\,\cite{mun} reported that the observed dynamical phase-diagram can be quantitatively explained by the theory based on the dynamical Gutzwiller approximation which was developed by E. Altman {\it et al.}\,\cite{altman} assuming the periodic boundary condition. We also note that energetic instabilities like the Landau instability do not occur in our calculation because we keep the total energy in the system constant during time evolution. Therefore, there is no dissipation, like the phonon in the Landau instability, discharging energy to a heat bath such as external environment or a thermal component in a system. This situation is consistent with the experiments that observed the dynamical instability. As shown in Ref.\,\cite{fallani}, energetic instabilities hardly appear in the experiments because the time scale of energetic instability is sufficiently longer than that of the dynamical instability at low temperatures where a thermal component is highly suppressed.

In this paper, we discuss the instability of a superfluid flow by introducing two characteristic momenta: a critical momentum $p_{c}$ and a decay momentum $p_{d}$.
$p_{c}$ corresponds to the critical momentum at which a superfluid flow starts to decay under the condition that its momentum is increased {\it adiabatically} from zero in an optical lattice. On the other hand, $p_{d}$ is the similar critical momentum when the momentum of a superfluid is increased at a certain acceleration rate $\alpha$ as in the real experiment situations. A superfluid flow actually starts to decay drastically at $p_{d}$ owing to the dynamical instability during time evolution governed by Eq.\,(\ref{eom}) based on the dynamical Gutzwiller method mentioned above. From these definitions, $p_{d}$ agrees with $p_{c}$ in the limit of $\alpha=0$. One can evaluate $p_{c}$ by the extrapolation using the values of $p_{d}$ at several acceleration rates, while we adopted the alternative approach mentioned below.

Here we briefly explain the way we calculate $p_{c}$ for the spinless case as a simple example \cite{polkov,asaoka}. The critical momentum $p_{c}$ is determined from the (non-dimensional) group velocity $v(p)=\rho(p)\sin(p)$ where $\rho(p)$ is the density of a {\it steady} superfluid flowing with momentum $p$. The periodicity of $v(p)$ reflects the structure of the lowest Bloch band in an optical lattice. In the framework of the Gutzwiller approximation, the density $\rho(p)$ is equivalent to the condensate fraction $n_{k=p}=\langle \hat a_{k=p}^{\dagger} \hat a_{k=p}\rangle=|\langle \hat a_{k=p}\rangle|^2$ defined as the population of the state with momentum $p$, where  $k$ is the quasi-momentum of a condensate in an optical lattice. The condensate fraction $n_{k=p} (\propto t'/U_0) $ is a monotonically decreasing function of $p$ according to the effective hopping amplitude $t'$ given by $t'=t(d+\cos p -1)/d$ where $d$ corresponds to the dimension of the system. Consequently, the group velocity $v(p)$ has a maximum at a certain momentum $p=p_{c} (< \pi/2)$ as $p$ is increased. Beyond this $p_c$, the effective mass, which is the inverse of the hopping amplitude in the tight binding model, becomes negative and then the sound velocity for the BHM becomes complex due to the formula $c_{s}=\frac{1}{\hbar}\sqrt{\frac{2t\rho}{\kappa}}$ based on the Gutzwiller approximation \cite{krutitsky}, where $\rho$ is the superfluid density and $\kappa\equiv\frac{\partial\langle n\rangle}{\partial\mu}$ is the compressibility. A superfluid flow is unstable above $p_{c}$ on such a mathematical background. 
Finally, the critical momentum $p_{c}$ is obtained self-consistently under the condition that the group velocity achieves its maximum value:
\begin{eqnarray}
p_{c}=\arctan\left(-\frac{n_{k=p=p_{c}}}{\left(\frac{dn_{k=p}}{dp}\right)_{p=p_{c}}}\right).
\label{pc}
\end{eqnarray}
Note that this equation is also applicable to the $S=1$ BHM.
We determine the phase boundary of the dynamical instability using this $p_{c}$ to remove the influence of momentum acceleration rate $\alpha$, while the previous work employed $p_{d}$ \cite{altman}.

\section{Results and Discussion}

\subsection{Dynamics of superfluid flow}

\begin{figure}
\includegraphics[clip,width=7.0cm]{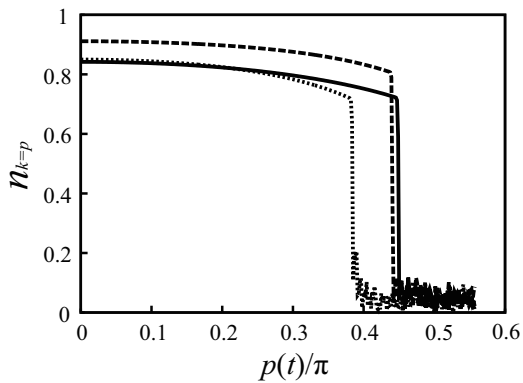}
\caption{
Time evolution of condensate fraction $n_{k=p}$ corresponding to three different cases: the $S=1$ Bose-Hubbard model (BHM) with $U_{2}/U_{0}=0.3$ (solid line) and $U_{2}/U_{0}=-0.3$ (dashed line), and the $S=0$ BHM (dotted line). The momentum $p$ is increased almost adiabatically in proportion to time $t$ such that $p(t) =0.005t$. We employ the system parameters $U_{0}=10$ and $n=1$. The decay momenta are correspondingly $p_{d} = 0.45$ for $U_{2}/U_{0}=0.3$ and 0.44 for $U_{2}/U_{0}=-0.3$ in the $S=1$ BHM, and 0.38 in the $S=0$ BHM.
}\label{nkp}
\end{figure}

Figure \ref{nkp} shows the time-evolution of condensate fractions $n_{k=p}$ for the spin-1 ($S=1$) and spinless  ($S=0$) BHM with filling $n=1$ and on-site repulsion $U_{0}=10$, as functions of increasing momentum in time where  $p(t)=0.005t$. In the $S=1$ BHM, we set $U_{2}/U_{0}=-0.3$ (ferromagnetic) and $U_{2}/U_{0}=0.3$ (antiferromagnetic). We choose the ground state as the initial state at time $t=0$ for each $U_{0}$ and $U_{2}$. The total $S_{z} (=\sum_{i} S_{i,z})$ in the system is conserved during the time evolution. 
From Fig.\,\ref{nkp}, $n_{k=p}$ gradually decreases as the momentum $p(t)$ is increased, and then suddenly decays owing to the dynamical instability. 
The decay momenta are correspondingly $p_{d} = 0.44$ for $U_{2}/U_{0}=-0.3$ and 0.45 for $U_{2}/U_{0}=0.3$ in the $S=1$ BHM, and 0.38 in the $S=0$ BHM. 
We find that the $S=1$ condensate persists to a larger $p(t)$ than the $S=0$ condensate, indicating some influences of the spin-dependent interaction included in the Hamiltonian Eq.\,(\ref{hamil}). 
Interestingly, the initial condensate fraction at $p(0) = 0$ in the $S=1$ BHM with $U_{2}/U_{0}=0.3$  is almost the same as that in the $S=0$ model. Moreover, even in the $S=1$ BHM, $p_{d}$ for $U_{2}/U_{0}=0.3$ is slightly larger than that for $U_{2}/U_{0}=-0.3$, while the condensate fraction around $p=p_{d}$ for $U_{2}/U_{0}=0.3$ is apparently smaller than that for $U_{2}/U_{0}=-0.3$.
These results suggest that the amplitude of the condensate fraction does not solely determine $p_{d}$, which is very consistent with the fact that the derivative $\frac{dn_{k=p}}{dp}$ is also included in Eq.\,(\ref{pc}) for determining $p_{c}$.

As we briefly mentioned in the previous section, the decay momentum $p_{d}$ inevitably becomes larger than the critical momentum $p_{c}$ of Eq.\,(\ref{pc}) when the system parameters are equal, i.e., the same interaction strength, filling, and lattice geometry. Our previous work in Ref.\,\cite{asaoka} showed that a superfluid can flow stably beyond the critical momentum $p_{c}$ until the unstable mode that causes dynamical instability fully grows.
This retardation effect always exists as long as a finite acceleration of a condensate exists in calculations or experiments.
We confirmed in Fig.\,\ref{nkp} that $p_{d}$ approaches $p_{c}$ in the both BHMs using a smaller coefficient $\alpha \,(<0.005)$ for $p(t)=\alpha t$.

%

\begin{figure}
\includegraphics[clip,width=9.0cm]{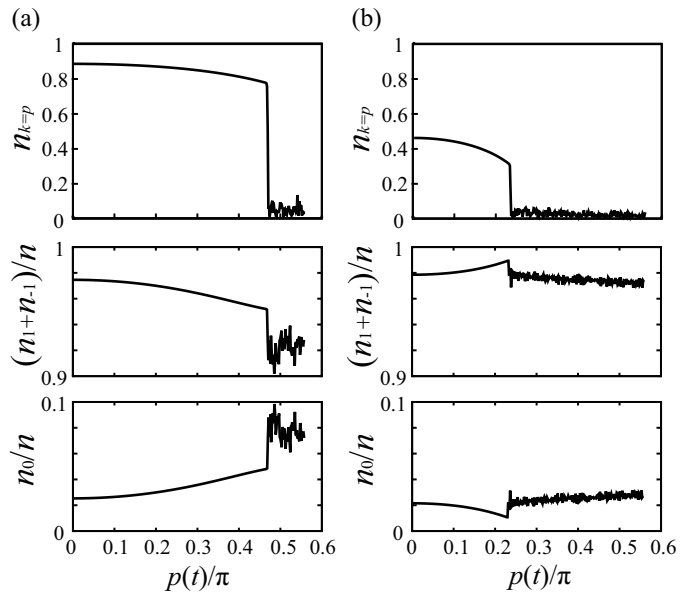}
\caption{
Time evolution of condensate fraction $n_{k=p}$ and population of each spin component $n_{\gamma}/n$ ($\gamma=0,\pm1$) in the $S=1$ BHM with antiferromagnetic interaction $U_{2}/U_{0}=0.3$: (a) $U_{0}/U_{0c}=0.2$ and (b) $U_{0}/U_{0c}=0.8$. Here $U_{0c}(=37.9)$ is the repulsive interaction strength at the Mott-insulator transition point. We set the filling at $n=1$.
}\label{time}

\end{figure}

 
Next we discuss the role played by the spin mixing processes during the time evolution in the $S=1 $ BHM, which is governed by the third and the fourth terms on the right-hand side of Eq.\,(\ref{eom}) in Appendix A. We focus on the antiferromagnetic case with $U_{2}/U_0=0.3$, in which the spin degrees of freedom are unfrozen. 
In our calculations, all particles are in the $S_{z}=1$ state and spins are completely frozen in the ferromagnetic case of $U_{2}<0$.  
Figure \ref{time} shows the time-evolution of the condensate fraction $n_{k=p}$ and the population of each spin component $n_{\gamma}/n\, (\gamma = 0, \pm1)$ for two interaction strengths: (a) $U_{0}/U_{0c} = 0.2$ and (b) $U_{0}/U_{0c} = 0.8$.  Here $U_{0c}$ denotes the critical interaction strength at the Mott-insulator transition point in the $S=1$ BHM and $U_{0c}=37.9$ for $U_{2}/U_0=0.3$.
Note that in Fig.\,\ref{time} both $n_{1}$ and $n_{-1}$ are always equal owing to the initial state we choose and the conservation of total $S_{z}$.
For $U_{0}/U_{0c}=0.2$ shown in Fig.\,\ref{time}\,(a), the populations of the $S_{z}=\pm1$ states gradually decrease and that of the $S_{z}=0$ state increases with increasing momentum, and finally all the spin components mix chaotically, which is accompanied by the decay of the superfluid flow.
We also find the similar chaotic mixing of the spin components for $U_{0}/U_{0c}=0.8$ in Fig.\,\ref{time}\,(b) after the populations of the $S_{z}=\pm1$ states have slightly increased and that of the $S_{z}=0$ state decreased. However, the variation in spin populations is very small during the time evolution, suggesting that the  spins are almost frozen in this case. We can naturally understand these results by noting that the third and fourth terms in Eq.\,(\ref{eom}) make a greater contribution to spin-mixing in the region where $U_{0}$ is sufficiently small and the amplitude of the Gutzwiller parameters $|f_i(n_{i}\geq 2)|^{2}$ becomes larger.
%

\subsection{Phase diagram at unit filling}

\begin{figure}
\includegraphics[clip,width=7.0cm]{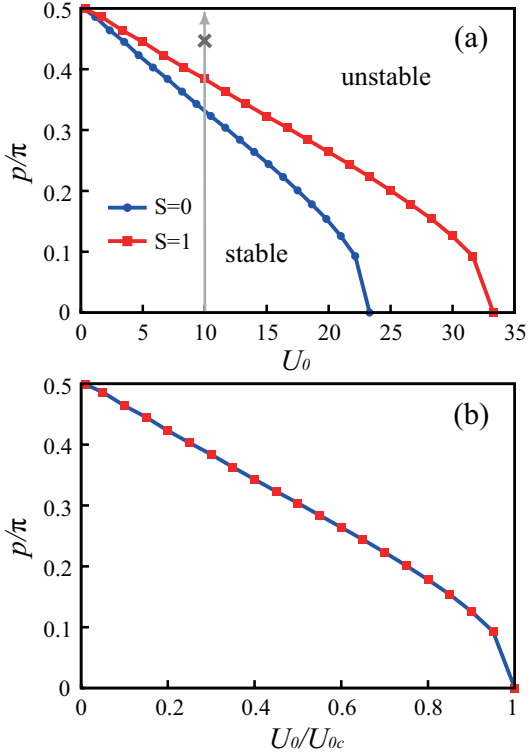}
\caption{
Dynamical phase diagrams of the $S=1$ BHM with ferromagnetic interaction $U_{2}/U_{0}=-0.3$ and the $S=0$ BHM as a function of (a) $U_{0}$ and (b) $U_{0}/U_{0c}$. 
A superfluid becomes dynamically unstable in the region above the phase boundary. The arrow in (a) corresponds to the horizontal axis in Fig.\,\ref{nkp}, and the cross indicates the decay momentum of the dashed line in Fig.\,\ref{nkp}. The phase boundaries of both models in (b) are completely identical.
We set the filling at $n=1$.
}\label{ferro}
\end{figure}

In this subsection, we discuss how the spin mixing processes in the $S=1$ BHM affect the critical momentum $p_c$ by focusing on the simple case of unit filling (i.e., $n=1$).
Figure\,\ref{ferro}\,(a) shows the dynamical phase diagram of the $S=1$ BHM with ferromagnetic interaction $U_{2}/U_{0}=-0.3$ along with the results of the $S=0$ BHM. 
Each line represents the critical momentum $p_c$ as a function of interaction strength $U_0$ and corresponds to the phase boundary that separates the stable and unstable phases. 
The dynamical instability occurs in the upper unstable region. 
Note that these phase boundaries are determined via the maximum of group velocity as is explained in Sec.\,\ref{sec:MM}.
Figure\,\ref{ferro}\,(a) shows that the critical momentum of the dynamical instability changes smoothly from $p_c=\pi/2$ at $U_{0}=0$ to $p_c=0$ at $U_0=U_{0c}$ (i.e., the interaction strength at the MI transition point in the thermal equilibrium). The critical interactions are $U_{0c}=33.3$ for the $S=1$ BHM and $U_{0c}=23.3$ for the $S=0$ BHM.
The cross at about $p/\pi=0.44$ in Fig.\,\ref{ferro} (a) on the dashed vertical line at $U_{0}=10$ represents the decay momentum of a spin-1 superfluid flow, $p_{d}$, seen in Fig.\,\ref{nkp}. The apparent discrepancy between this point and the phase boundary is due to the retardation effect in the dynamical instability as explained in relation to Fig.\,\ref{nkp}.

We examine this dynamical phase diagram in more detail. In our calculations, all spin-1 particles with ferromagnetic interaction stay in the $S_{z}=1$ state, which makes the situation relatively simple. Therefore the spin dependent $U_{2}$ term in the Hamiltonian Eq.\,(\ref{hamil}) becomes
\begin{eqnarray}
\frac{U_{2}}{2}\sum_{i}\hat{n}_{i}(\hat{n}_{i}-1)        .
\label{ferro_U2}
\end{eqnarray}
Since this form is equal to the spinless $U_{0}$ term in the Hamiltonian, the $U_{2}$ term gives just the shift in the $U_{0}$ value (i.e., $ U_0 \to U_{0}+U_{2}$). 
In the present case of $U_{2}=-0.3U_{0}$, $U_{0}$ is effectively reduced to $0.7U_{0}$. As is shown in Fig.\,\ref{ferro}\,(b), both phase boundaries overlap completely when $U_{0}$ is normalized with each $U_{0c}$. Thus the dynamical instability in the ferromagnetic case is essentially equivalent to that in the spinless case.
%

\begin{figure}
\includegraphics[clip,width=7.0cm]{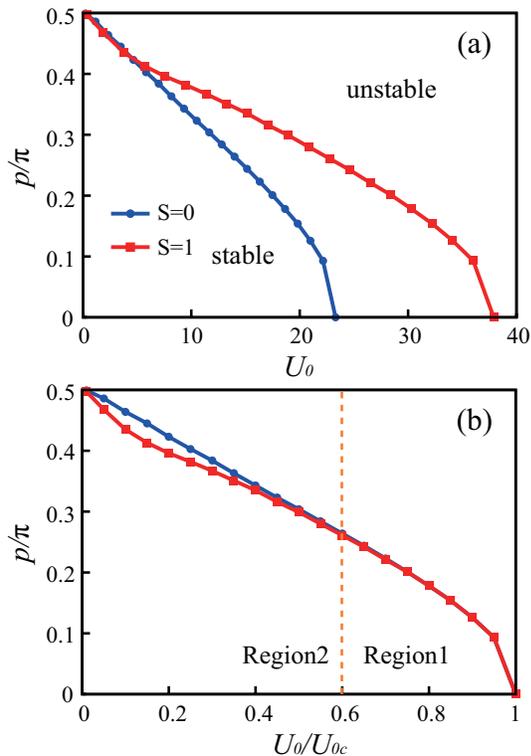}
\caption{
Dynamical phase diagrams of the $S=1$ BHM with antiferromagnetic interaction $U_{2}/U_{0}=0.3$ and the $S=0$ BHM as a function of (a) $U_{0}$ and (b) $U_{0}/U_{0c}$. 
In (b), we divide the phase diagram into two regions: region 1 for $U_{0}/U_{0c} > 0.6$ and region 2 for $U_{0}/U_{0c} < 0.6$. We set the filling at $n=1$.
}\label{antiferro}
\end{figure}

Next, we discuss the antiferromagnetic case. Figure\,\ref{antiferro}\,(a) shows the dynamical phase diagram of the $S=1$ BHM with antiferromagnetic interaction $U_{2}/U_{0}=0.3$ along with the results of the $S=0$ BHM. 
We find that the two phase boundaries are very close together for $U_{0} \lesssim 5$ (the $S=1$ boundary is slightly below), and gradually diverge for $U_{0} \gtrsim 5$.
This divergence of the phase boundaries for $U_{0} \gtrsim 5$ basically originates from the difference in the Mott-transition points between the $S=0$ and 1 BHMs.
In the strongly correlated regime, the probability of double occupation $n_{j}=2$ at each site in the $S=1$ model is much larger than that in the $S=0$ model for the same interaction strength $U_{0}$ owing to the formation of the local spin singlet state $|n_{j},S_{j},S_{j,z}\rangle=|2,0,0\rangle$ [this formulation is defined by formula (23) in reference \cite{yama}], which has the energy gain $-2U_{2}$ in the $U_{2}$ term. This enhanced number fluctuation leads to a larger critical interaction strength at the Mott-transition point $U_{0c} $ in the $S=1$ BHM with unit filling. Correspondingly, the stable area of the $S=1$ model in the phase diagram grows compared with that of the $S=0$ system.

In Fig.\,\ref{antiferro}\,(b), we present the same phase diagram as a function of normalized interaction strength $U_0/U_{0c}$. 
The phase boundaries of the two models are very close together for $U_{0}/U_{0c} \gtrsim 0.6$, gradually diverge for $U_{0}/U_{0c} \lesssim 0.6$, and finally reach $p_c=\pi/2$ at $U_{0}/U_{0c}=0$.
In contrast to the ferromagnetic case shown in Fig.\,\ref{ferro}\,(b), the different phase boundaries for $U_{0}/U_{0c} \lesssim 0.6$ clearly reflect that 
the spin mixing processes included in the $S=1$ BHM [i.e., the last two terms on the right hand side of Eq.\,(\ref{sp})] play a role in the antiferromagnetic case and influence the critical momentum of the dynamical instability. 
We examine this effect by dividing the phase diagram into two regions: a strongly correlated regime in which the phase boundaries are close together (region 1 for $U_{0}/U_{0c} > 0.6$) and another regime in which the boundaries diverge (region 2 for $U_{0}/U_{0c} < 0.6$).

\begin{figure}
\includegraphics[clip,width=7.0cm]{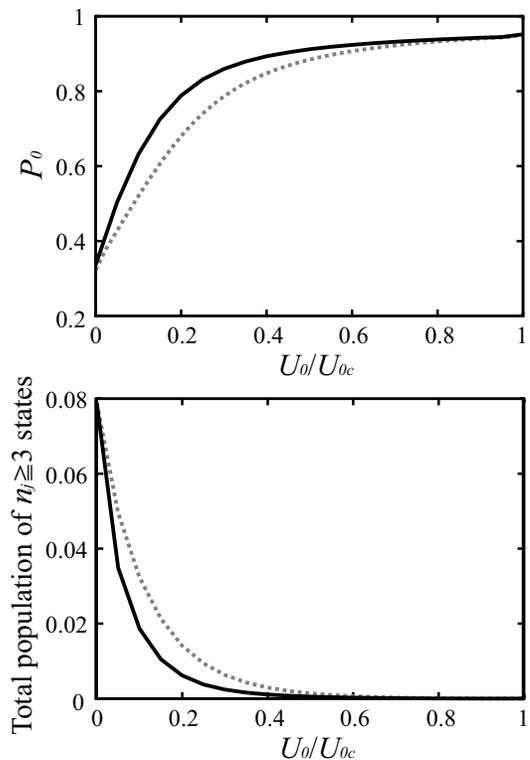}
\caption{
(a): Populations of the local spin singlet state $|S_{j}=0\rangle$ at $p=0$ (dashed line) and $p=p_{c}$ (solid line). (b): Total population of $n_{j}\geq 3$ states within our truncated Fock space at $p=0$ (dashed line) and $p=p_{c}$ (solid line). We employ $U_{2}/U_{0}=0.3$ and $n=1$ as in Fig.\,\ref{antiferro}.
}\label{spin_fra}
\end{figure}

We begin by explaining region 1 in which the phase boundaries overlap.
For this purpose, we assume a system where the maximum number of particles per site is $n_{\text{max}}=2$ because the number fluctuations are greatly suppressed in a strongly correlated regime and the probability of $n_{j}\geq 3$ states is negligible.
By noting that the $U_2$ term vanishes for the $n_{j}=1$ state, only the states, $|n_{j},S_{j},S_{j,z}\rangle = |2,0,0\rangle$ and $|2,2,\eta \rangle$ ($\eta=0,\pm1,\pm2$), have non-zero energies corresponding to the $U_{2}$ term: $-2U_{2}$ for $|2,0,0\rangle$ and $U_{2}$ for $|2,2,\eta \rangle$. The $|2,2,\eta \rangle$ states are degenerate under the current condition without a magnetic field. For simplicity, we define the local spin states as $|S_{j}=0\rangle\equiv|2,0,0\rangle$ and $|S_{j}=2\rangle\equiv\frac{1}{\sqrt{5}}\sum_{\eta}|2,2,\eta \rangle$.
The population of the local singlet state $|S_{j}=0\rangle$ included in the $n_{j}=2$ state is evaluated via the Gutzwiller parameters: 
\begin{eqnarray}
\nonumber 
P_{0}&=&\frac{\langle \Psi_{\mathrm G} |S_{j}=0\rangle \langle S_{j}=0| \Psi_{\mathrm G}\rangle}
{\langle \Psi_{\mathrm G} |S_{j}=0\rangle \langle S_{j}=0| \Psi_{\mathrm G}\rangle+\langle \Psi_{\mathrm G} |S_{j}=2\rangle \langle S_{j}=2| \Psi_{\mathrm G}\rangle},\\
\nonumber &=& \frac{|\langle\Psi_{\mathrm G} |S_{j}=0\rangle|^{2}}{|\langle\Psi_{\mathrm G}|S_{j}=0\rangle|^{2}+|\langle\Psi_{\mathrm G}|S_{j}=2\rangle|^{2}},\\
&=&\frac{\frac{1}{3}|f(0,2,0)-\sqrt{2}f(1,0,1)|^{2}}{\sum_{n_{i,1}+n_{i,0}+n_{i,-1}=2}|f(n_{i,1},n_{i,0},n_{i,-1})|^{2}}.
\end{eqnarray}
Furthermore, the population of the $|S_{j}=2\rangle$ state is given by $P_{2}=1-P_{0}$. 
In Fig.\,\ref{spin_fra}\,(a), we show $P_0$  at both $p=0$ (solid line) and $p=p_c$ (dashed line) as a function of $U_0$. 
Note that, in region 1, there is hardly any change in $P_0$ or $P_2$ irrespective of the interaction strength and the superfluid momentum, reflecting the fact that the spin state becomes stationary in this region. This result is consistent with the slight spin variation seen in Fig.\,\ref{time} (b). The spin mixing process does not occur in region 1, and consequently the spin dependent $U_{2}$ term causes only the shift in $U_{0}$ as in the ferromagnetic case. We can thus understand that the phase boundaries of both the $S=0$ and 1 BHMs become identical when $U_{0}$ is normalized by the corresponding $U_{0c}$ as in Fig.\,\ref{antiferro} (b).

On the other hand, in region 2, the spin configurations become complex because the population of the $n_{i}\geq 3$ states increases, and the spin mixing in the $U_{2}$ term plays a role (Fig.\,\ref{spin_fra} (b)).
Here we examine how the spin degrees of freedom influence the value of the critical momentum $p_{c}$ and discuss the origin of the divergence of the phase boundaries in region 2 seen in Fig.\,\ref{antiferro} (b). 
First, it follows from Eq.\,(\ref{pc}) that $p_{c}$ is monotonically proportional to $n_{k=p=p_{c}}/\left|\left(\frac{dn_{k=p}}{dp}\right)_{p=p_{c}}\right|$ because $\frac{dn_{k=p}}{dp} < 0$ for a stable superfluid flow and $n_{k=p} > 0$.
Furthermore, in a weakly correlated regime, the condensate fractions in both the $S=0$ and 1 systems are sufficiently large and equally close to 1. The difference between the $p_{c}$ values of these two systems is therefore determined largely by the $|dn_{k=p}/dp|$ factor in the denominator of the above relation. 

Next we explain the influence of spins on this factor of $|dn_{k=p}/dp|$. It is generally known that the effective hopping amplitude of a condensate carrying the momentum $p$ becomes $t'=t \cos(p)$. The increment in momentum hence diminishes the condensate fraction $n_{k=p}$ as is shown in Fig.\,\ref{nkp}, which simultaneously reduces the number fluctuations of a condensate. In the $S=1$ BHM, this effect becomes more prominent thanks to the $U_{2}$ term in the Hamiltonian. Figure\,\ref{spin_fra} shows that the population of the $|S_{i}=0\rangle$ state increases with increasing momentum in the weakly correlated regime, while the population of $n_{i}\geq 3$ states decreases. This suggests that the number fluctuations in the $S=1$ system are further suppressed in order to gain an energy of $-2U_{2}$ in the $U_{2}$ term. Therefore, in the weakly correlated regime, $|dn_{k=p}/dp|$ when $S=1$ is generally larger than that when $S=0$. We can confirm this fact numerically from our present result: $\delta n_{k=p}=n_{k=p=0}-n_{k=p=p_{c}}$ is 0.045 for $S=1$ with $U_{2}/U_{0}=0.3$, which is nearly two times higher than 0.025 when $S=0$, for $U_{0}/U_{c}=0.1$ in both cases. Returning to Fig.\,\ref{antiferro} (a), the difference between the phase boundaries of $S=1$ and 0 is very small in the weakly correlated regime owing to the small $U_{2}$ values there. However, the normalized phase diagram in Fig.\,\ref{antiferro} (b) successfully extracts the existence of the spin effect on the dynamical instability of a superfluid flow. 

\subsection{Phase diagrams at other fillings}

\subsubsection{Commensurate case}

It is generally known that the SF-MI transition in the BHM strongly depends on fillings (i.e., the average number of particles per site). Specifically, in the $S=1$ BHM with antiferromagnetic interactions, the critical interaction strength at the transition $U_{0c}$ shows a clear dependence on the parity of fillings: $U_{0c}$ at odd fillings is larger than that in the $S=0$ BHM system, while it becomes smaller at even fillings \cite{tsuchiya,kimura}.
This property is easily understood from the fact that the formation of the local singlet state to gain an energy of $-2U_{2}$ in the $U_{2}$ term in the Hamiltonian Eq.\,(\ref{hamil}) enhances (suppresses) the density fluctuations at odd (even) fillings. 
Here we discuss how the parity affects the dynamical instability in the $S=1$ BHM.

\begin{figure}
\includegraphics[clip,width=8.0cm]{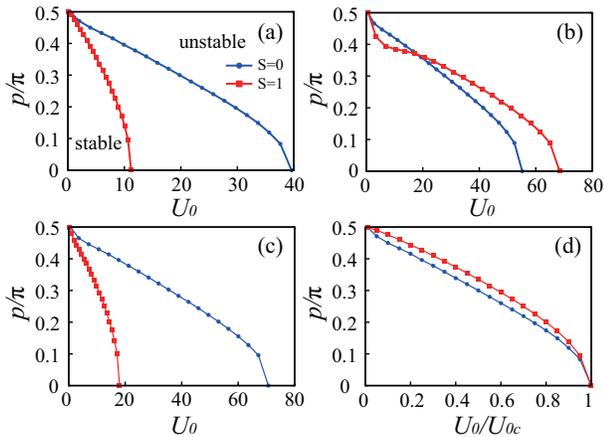}
\caption{
Dynamical phase diagrams in the $S=1$ and $S=0$ BHMs for three different fillings: (a) $n=2$, (b) $n=3$, and (c) $n=4$. In (d), the results of (a) are presented as a function of $U_{0}/U_{0c}$. $U_{2}/U_{0}$ is fixed at 0.3 in all cases for the $S=1$ model.
Critical interaction strengths are $U_{0c}=39.6 \,(n=2)$, $55.2 \,(n=3)$, and $70.7 \,(n=4)$ for $S=0$, and $U_{0c}=11.2 \,(n=2)$, $68.45 \,(n=3)$, and $18.0 \,(n=4)$ for $S=1$.
}\label{diag3}
\end{figure}

In Fig.\,\ref{diag3} (a)-(c), the dynamical phase diagrams of the $S=1$ BHM for $U_{2}/U_{0}=0.3$ are given for the several different fillings (i.e., $n=$2, 3, and 4) along with the results of the $S=0$ model. 
From these figures and Fig.\,\ref{antiferro} (a), we find that the influence of the parity clearly appears in the dynamical phase diagrams. The stable areas of the $S=1$ model basically grow (shrink) at even (odd) fillings compared with the $S=0$ model, which reflects the corresponding increase (decrease) of $U_{0c}$.
With $n=3$ shown in Fig.\,\ref{diag3} (b), however, the stable area clearly decreases in the weakly correlated regime. As we pointed out for unit filling in the previous subsection, the unfrozen spins that prefer to form the local singlet state greatly suppress the density fluctuations and make the superfluid flow unstable in the weakly correlated regime. We have confirmed this effect more clearly for $n=3$ filling. This result indicates that the spin mixing process has a greater influence at larger fillings.

In Fig.\,\ref{diag3}\,(d), we provide a dynamical phase diagram for $n=2$ filling as a function of the normalized  interaction $U_{0}/U_{0c}$. The $S=1$ phase boundary is located above the $S=0$ curve over the entire interaction range, which is in contrast to the phase diagram for unit filling shown in Fig.\,\ref{antiferro}\,(b).
This suggests that the unfrozen spins, which prefer the local singlet states, stabilize the superfluid flow. 
We have also found this tendency with $n=4$ filling as a characteristic of the dynamical instability at even fillings. 
%

\subsubsection{Incommensurate case}


In a system with an incommensurate filling, the SF-MI transition does not occur because of the extra particles deviating from the commensurate filling. 
Polkovnikov {\it et al.} calculated the dynamical phase diagram of the $S=0$ BHM with incommensurate fillings based on the Gutzwiller approximation in Ref.\,\cite{polkov}.
They clarified that the critical momentum $p_{c}$ has a minimum value at a certain interaction strength and then asymptotically approaches $\pi/2$ with increase in interaction strength. 
The superfluidity of the extra particles becomes highly robust in the strongly interacting regime, i.e., the superfluidity recovers owing to the repulsive interaction.
The minimum of $p_c$ in the dynamical phase diagram therefore represents the crossover between weakly and strongly interacting regimes.
Here we analyze this tendency in the $S=1$ BHM with the antiferromagnetic interaction $U_{2}/U_0=0.3$.

\begin{figure}
\includegraphics[clip,width=8.0cm]{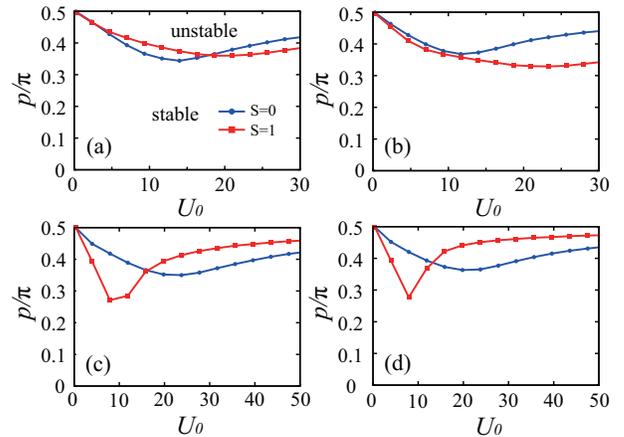}
\caption{
Dynamical phase diagram for incommensurate fillings: (a) $n=0.8$, (b) $n=1.2$, (c) $n=1.8$, and (d) $n=2.2$. We employ $U_{2}/U_{0}=0.3$ in all cases for the $S=1$ model.
}\label{incomme}
\end{figure}

Figure\,\ref{incomme} shows the dynamical phase diagrams of the $S=1$ BHM with the filling factors deviating slightly from $n=1$ and 2, along with diagrams of the $S=0$ BHM. 
From these figures, the dynamical phase diagrams of the $S=1$ BHM at incommensurate fillings agree qualitatively  with those of the spinless $S=0$ model.
However, we still find the influence of the parity, which we have seen in the phase boundaries for the commensurate cases. 
In Fig.\,\ref{incomme}\,(a) and (b) where the fillings are close to $n=1$, the critical momentum reaches its minimum value at a larger interaction strength in the $S=1$ BHM in comparison with the $S=0$ results. On the other hand, in Fig.\,\ref{incomme}\,(c) and (d) where the fillings are close to $n=2$, the minimum point for $S=1$ is apparently smaller than that for $S=0$.
This behavior can be roughly understood in terms of whether the formation of the local spin singlet state conceals or accentuates the extra particles. As mentioned in the commensurate case, the formation of the local singlet state enhances (suppresses) the density fluctuation of a condensate with the odd (even) fillings. The intense density fluctuation of a condensate conceals the effect of the extra particles on the left of the minimum points while the suppressed density fluctuation accentuates the extra particles on the right side. Therefore, with the fillings close to $n=1$, the extra particles are more concealed due to the formation of the local spin singlet state, and the minimum point for $S=1$ slides to the right compared to that of $S=0$, while the extra particles are more accentuated due to the formation of the local spin singlet state and the minimum point for $S=1$ slides to the left compared with that of $S=0$ with the fillings close to $n=2$.
Furthermore, we see that there is particle-hole symmetry in the dynamical phase diagrams with incommensurate fillings by noting the consistency between the $n=0.8$ and 1.2 results, and also between the $n=1.8$ and 2.2 results.
%

\subsection{Density modulation}


Finally, we discuss the density and spin modulation associated with the dynamical instability.
A spinless superfluid flow in an optical lattice exhibits a density modulation as a precursor to dynamical instability \cite{fallani,sarlo}. This is a manifestation of unstable collective excitation modes as a seed of the dynamical instability, and it depends strongly on the interaction strength or the acceleration rate of the condensate momentum \cite{asaoka}.
This collective excitation, which involves a lot of physical information, is significant in terms of understanding the dynamical instability.
Here we examine whether the occurrence of spin modulation is associated with the dynamical instability of the spin-1 condensate in the $S=1$ BHM.
We again assume an antiferromagnetic system with a spin-dependent interaction strength of $U_{2}/U_{0}=0.3$.
%
\begin{figure}
\includegraphics[clip,width=7.0cm]{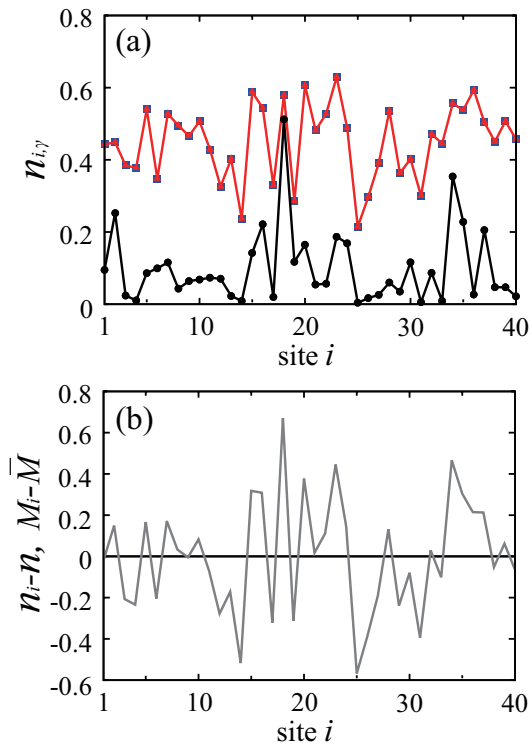}
\caption{
Density modulation associated with the dynamical instability with no magnetic field.
(a): Density distributions of $S_{z}=1$ (triangles), $S_{z}=0$ (circles), and $S_{z}=-1$ (squares) components over the lattice sites. The density distributions of $S_{z}=\pm1$ are fully identical. 
(b): Deviation of densities from the mean values: particle number (light line) and magnetization (dark line).
The parameters are fixed at $n=1$, $U_{0}=10$, and $U_{2}/U_{0}=0.3$. These results correspond to the point at $p /\pi=0.46$ after the decay at $p_{d}/\pi=0.45$ on the solid line in Fig.\,\ref{nkp}.
}\label{modu_nomag}
\end{figure}

Figure\,\ref{modu_nomag} (a) shows the density distributions of $S_{z}=0,\pm1$ components after the dynamical decay of a condensate. The density modulation develops sufficiently at this momentum. We find that the density modulations of the $S_{z}=\pm1$ components develop in unison, and that of the $S_{z}=0$ component develops independent of those modulations. This result indicates that small spatial fluctuations in a condensate grow independently in $S_{z}=\pm1$ components and the $S_{z}=0$ component. This is because the components of $S_{z}=\pm1$ are equivalent in Eq.\,(\ref{eom}) within the mean-field approximation where the density modulations of $S_{z}=\pm1$ components develop in unison.
Figure\,\ref{modu_nomag} (b) shows the total density distribution and magnetization distribution at the same momentum.
There is no spin modulation of the $S_{z}$ components while the total density modulation develops intensely. This reflects the consistent development of the modulations of the $S_{z}=\pm1$ components. Therefore, the spin modulation occurs only within the $xy$ plane.
%
\begin{figure}
\includegraphics[clip,width=7.0cm]{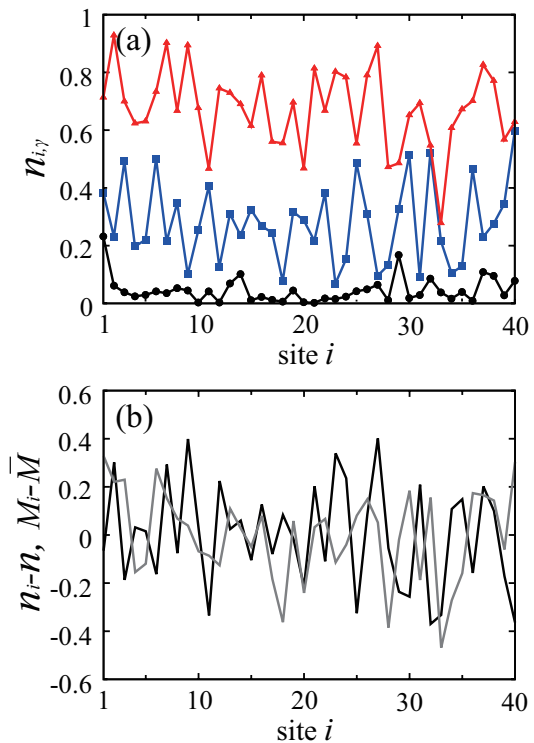}
\caption{
Density modulation associated with dynamical instability under a uniform magnetic field.
(a): Density distributions of $S_{z}=1$ (triangles), $S_{z}=0$ (circles), and $S_{z}=-1$ (squares) components over the lattice sites. 
(b): Deviation of densities from the mean values:  particle number (light line) and magnetization (dark line).
The parameters are $n=1$, $U_{0}=10$, and $U_{2}/U_{0}=0.3$ as in Fig\,\ref{modu_nomag}.
A uniform magnetic field is used to adjust the initial populations of the spin components to $n_{1}:n_{-1}\sim 7:3$ under the condition of $p=0$.
These results correspond to the point at $p/\pi=0.44$ after the decay at $p_{d}/\pi=0.427$ as the momentum is increased such that $p(t)=0.005t$.
}\label{modu_mag}
\end{figure}

Next, let us examine the spin modulation in a system with a uniform magnetic field in the $z$ direction.
Here, the $S_{z}=\pm1$ populations become imbalanced and we adjust them to $n_{1}:n_{-1}\sim 7:3$.
Figure\,\ref{modu_mag} (a) shows the density distributions of $S_{z}=0,\pm1$ components after the dynamical decay of a condensate. A magnetic field is applied to the system only at initial state $p=0$, but the initial state is stable since the total $S_{z}$ in the system is conserved.
A significant difference from non-magnetic case is that the density modulations of the $S_{z}=\pm1$ components develop independently, namely, small spatial fluctuations in a condensate grow independently in $S_{z}=1$ and $S_{z}=-1$ components.
This result indicates that $S_{z}=\pm1$ components decay independently only if there is difference between the component populations.
As a result, the spin modulation of $S_{z}$ components occurs as shown in Fig.\,\ref{modu_mag} (b). 

\section{Summary}

In this study, we analyzed the dynamical instability of a superfluid flow in the $S=1$ BHM using the Gutzwiller approximations. Time evolutional calculations revealed that the superfluid flow of the spin-1 condensate decays at a different critical momentum from the spinless model when the interaction strength is the same, which is due to  spin-dependent interactions. Furthermore, we obtained the dynamical phase diagrams of both the $S=1$ and spinless $S=0$ BHMs and discussed their differences. With a ferromagnetic interaction $U_{2}<0$, the phase diagram of the spin-polarized $S=1$ BHM becomes essentially the same as the diagram of the spinless BHM because we can appropriately renormalize the interactions. On the other hand, with an antiferromagnetic interaction $U_{2}>0$, the dynamical phase diagrams of the $S=1$ BHM differ fundamentally from the spinless model and shed light on the influence of the spin mixing process between the $S=1$ bosons. We discussed in detail the important role of the formation of the local singlet state in the dynamical instability and the SF-MI transition in the $S=1$ BHM. 
Our systematical study also showed that the phase diagram strongly depends on the average number of particles per site, in particular, the even-odd parity.
We finally discussed the density modulation and the spin modulation associated with the dynamical instability. We found that the anisotropy of the spin modulation depends on whether or not a uniform magnetic field is present. This suggests that the spin modulation is highly sensitive to the imbalance in the spin components generated by a uniform magnetic field.

\begin{acknowledgments}
Some of the numerical computations were carried out at the Yukawa Institute Computer Facility and at the Cyberscience Center, Tohoku University. This work was supported by Japan Society for the Promotion of Science KAKENHI Grant No. 25287104.
\end{acknowledgments}

\appendix
\section{The equations of spin-1 Gutzwiller parameters}

The equations of motion for the Gutzwiller parameters in the $S=1$ BHM are given as

\begin{widetext}
\begin{eqnarray}
\nonumber
i\dot{f}_{j}(n_{j,-1},n_{j,0}&&,n_{j,1}) = \frac{U_{0}}{2}n_{j}(n_{j}-1)f_{j}(n_{j,-1},n_{j,0},n_{j,1})
\nonumber \\ &&+\frac{U_{2}}{2}(n_{j,1}^2-2n_{j,1}n_{j,-1}+n_{j,-1}^2-n_{j,1}-n_{j,-1}+2n_{j,1}n_{j,0}+2n_{j,0}n_{j,-1})f_{j}(n_{j,-1},n_{j,0},n_{j,1})
\nonumber \\ &&+U_{2}\sqrt{n_{j,1}(n_{j,0}+1)(n_{j,0}+2)n_{j,-1}}f_{j}(n_{j,-1}-1,n_{j,0}+2,n_{j,1}-1)
\nonumber \\ &&+U_{2}\sqrt{(n_{j,1}+1)n_{j,0}(n_{j,0}-1)(n_{j,-1}+1)}f_{j}(n_{j,-1}+1,n_{j,0}-2,n_{j,1}+1)
\nonumber \\ &&-tz\left(\sqrt{n_{j,-1}}f_{j}(n_{j,-1}-1,n_{j,0},n_{j,1})\psi_{j,-1}+\sqrt{n_{j,-1}+1}f_{j}(n_{j,-1}+1,n_{j,0},n_{j,1})\psi _{j,-1}^{*}\right)
\nonumber \\ &&-tz\left(\sqrt{n_{j,0}}f_{j}(n_{j,-1},n_{j,0}-1,n_{j,1})\psi_{j,0}+\sqrt{n_{j,0}+1}f_{j}(n_{j,-1},n_{j,0}+1,n_{j,1})\psi _{j,0}^{*}\right)
\nonumber \\ &&-tz\left(\sqrt{n_{j,1}}f_{j}(n_{j,-1},n_{j,0},n_{j,1}-1)\psi_{j,1}+\sqrt{n_{j,1}+1}f_{j}(n_{j,-1},n_{j,0},n_{j,1}+1)\psi _{j,1}^{*}\right) ,
\nonumber \\
\label{eom}
\end{eqnarray}
where

\begin{eqnarray}
\nonumber \psi _{j,1} &=& \frac{1}{z}
\sum_{\substack{n_{j+1,0}\\n_{j+1,\pm1}}}\sqrt{n_{j+1,1}+1}f_{j+1}^{*}(n_{j+1,-1},n_{j+1,0},n_{j+1,1})f_{j+1}(n_{j+1,-1},n_{j+1,0},n_{j+1,1}+1)e^{ip(t)}
\nonumber  \\
&& + \left. \frac{1}{z}\sum_{\substack{n_{j-1,0}\\n_{j-1,\pm1}}}\sqrt{n_{j-1,1}+1}f_{j-1}^{*}(n_{j-1,-1},n_{j-1,0},n_{j-1,1})f_{j-1}(n_{j-1,-1},n_{j-1,0},n_{j-1,1}+1)e^{-ip(t)}\right.
\nonumber \\
&&+  \frac{1}{z}\sum_{\tau}\sum_{\substack{n_{j+\tau,0}\\n_{j+\tau,\pm1}}}\sqrt{n_{j+\tau,1}+1}
f_{j+\tau}^{*}(n_{j+\tau,-1},n_{j+\tau,0},n_{j+\tau,1})f_{j+\tau}(n_{j+\tau,-1},n_{j+\tau,0},n_{j+\tau,1}+1) .
\nonumber
\end{eqnarray}
\begin{eqnarray}
\nonumber \psi _{j,0} &=& \frac{1}{z}
\sum_{\substack{n_{j+1,0}\\n_{j+1,\pm1}}}\sqrt{n_{j+1,0}+1}f_{j+1}^{*}(n_{j+1,-1},n_{j+1,0},n_{j+1,1})f_{j+1}(n_{j+1,-1},n_{j+1,0}+1,n_{j+1,1})e^{ip(t)}
\nonumber  \\
&& + \left. \frac{1}{z}\sum_{\substack{n_{j-1,0}\\n_{j-1,\pm1}}}\sqrt{n_{j-1,0}+1}f_{j-1}^{*}(n_{j-1,-1},n_{j-1,0},n_{j-1,1})f_{j-1}(n_{j-1,-1},n_{j-1,0}+1,n_{j-1,1})e^{-ip(t)}\right.
\nonumber \\
&&+  \frac{1}{z}\sum_{\tau}\sum_{\substack{n_{j+\tau,0}\\n_{j+\tau,\pm1}}}\sqrt{n_{j+\tau,0}+1}
f_{j+\tau}^{*}(n_{j+\tau,-1},n_{j+\tau,0},n_{j+\tau,1})f_{j+\tau}(n_{j+\tau,-1},n_{j+\tau,0}+1,n_{j+\tau,1}) .
\nonumber
\end{eqnarray}
\begin{eqnarray}
\nonumber \psi _{j,-1} &=& \frac{1}{z}
\sum_{\substack{n_{j+1,0}\\n_{j+1,\pm1}}}\sqrt
{n_{j+1,-1}+1}f_{j+1}^{*}(n_{j+1,-1},n_{j+1,0},n_{j+1,1})f_{j+1}(n_{j+1,-1}+1,n_{j+1,0},n_{j+1,1})e^{ip(t)}
\nonumber  \\
&& + \left. \frac{1}{z}\sum_{\substack{n_{j-1,0}\\n_{j-1,\pm1}}}\sqrt{n_{j-1,-1}+1}f_{j-1}^{*}(n_{j-1,-1},n_{j-1,0},n_{j-1,1})f_{j-1}(n_{j-1,-1}+1,n_{j-1,0},n_{j-1,1})e^{-ip(t)}\right.
\nonumber \\
&&+  \frac{1}{z}\sum_{\tau}\sum_{\substack{n_{j+\tau,0}\\n_{j+\tau,\pm1}}}\sqrt{n_{j+\tau,-1}+1}
f_{j+\tau}^{*}(n_{j+\tau,-1},n_{j+\tau,0},n_{j+\tau,1})f_{j+\tau}(n_{j+\tau,-1}+1,n_{j+\tau,0},n_{j+\tau,1}) .
\nonumber \\
\label{eom2}
\end{eqnarray}
\end{widetext}

Here $j+1$, $j-1$, and $j+\tau$ denote $(j_{1}+1,j_{2})$, $(j_{1}-1,j_{2})$, and $(j_{1},j_{2}+\tau)$ respectively, where $j_{1}$ is the site index of the flow direction and $j_{2}$ is that of the orthogonal direction. The summation $\sum_{\tau}$ runs over the nearest neighbors of site $j$ in the orthogonal direction, and $z$ is the number of adjacent sites in the lattice.

\end{document}